\documentclass[
reprint,
notitlepage,
 amsmath,amssymb,
 aps,
prx,
]{revtex4-2}

\usepackage[T1]{fontenc} 
\usepackage{amsmath}
\usepackage{amsfonts}
\usepackage{amsthm}
\usepackage{mathtools}
\usepackage{xcolor}  
\usepackage{wrapfig}
\usepackage{graphicx}
\usepackage{xurl}
\newcommand{\us}{{\bf s}}
\newcommand{\uS}{{\bf S}}

\newcommand{\uf}{{\bf f}}

\newcommand{\uX}{{\bf X}}

\newcommand{\uF}{{\bf F}}
\newcommand{\uh}{{\bf h}}
\newcommand{\ux}{{\bf x}}

\newcommand{\uy}{{\bf y}}

\newcommand{\kb}{k_{\rm B}}

\newcommand{\uint}{U_{\rm int}}

\newcommand{\rd}{{\rm d}}

\usepackage[%
  colorlinks=true,
  allcolors=blue
]{hyperref}

\usepackage{natbib}
\usepackage{textcomp}

\begin{document}

\title[picdiff]{Stitching Molecular Worlds Together with Physics-Coupled Diffusion Models}
\author{Yanbin Wang$^{1}$, Yajie Cai$^{1}$, and Ming Chen$^{1,\dagger}$} \affiliation{$^{1}$Department of Chemistry, Purdue University, West Lafayette, Indiana 47907, USA}
\email{chen4116@purdue.edu}



\begin{abstract}
Complex chemical systems often contain multiple components or large molecules, giving rise to substantial chemical and conformational complexity. Modeling the structures of these systems is crucial for mechanistic understanding of chemical processes and rational design, but remains highly challenging for both conventional theoretical approaches and machine learning. 
Here, we introduce physical-interaction-coupled diffusion models (PICDiff), a framework that couples independently trained diffusion models for subsystems of a complex system through explicit bonded and nonbonded interactions during inference. 
PICDiff reduces the difficulty of applying generative models to chemically and conformationally complex systems by decomposing them into smaller subsystems that are less complex and more amenable to machine-learning-based modeling. 
Using peptide–polymer binding and polymer conformation sampling as examples, we demonstrate that PICDiff can quantitatively sample conformations and model the thermodynamics of complex chemical systems.
These results show that PICDiff provides a general and practical approach for modeling complex chemical systems by combining learned models of simpler molecular subsystems through physical interactions.
\end{abstract}

\maketitle

\section{INTRODUCTION}
Molecular systems often emerge from the interaction of components or fragments whose individual equilibrium ensembles are easier to sample than the full coupled system. Examples include biomolecules interacting with soft interfaces \cite{sarker2022hydration,wei2012molecular,kim2025all,pascolutti2016structure}, polymers assembled from shorter blocks\cite{le2019synthesis,liang2022rapid,bates1990block,li2025synergistic}, and biomolecular condensate\cite{chang2024peptide,yuan2023peptide}. In these systems, the relevant equilibrium behavior is not determined solely by the conformational statistics of each isolated component, but by the physical interactions that couple these components into a joint ensemble\cite{frenkel2001understanding,chandler1987introduction}. Directly sampling this coupled ensemble with molecular simulations can be computationally expensive, especially when slow conformational changes, interfacial rearrangements, or long-chain structural fluctuations are involved\cite{allison2020computational,henin2022enhanced,bernardi2015enhanced,prentiss2010energy,chen2015locating,knight2011multisite,laio2002escaping,maragliano2006temperature,abrams2008efficient}. These challenges raise a general question: can independently learned molecular ensembles be composed into a thermodynamically meaningful interacting ensemble without exhaustively sampling the full coupled system?

Machine-learning generative models provide a promising route for molecular conformational sampling because they can learn high-dimensional structural distributions from simulation or structural data and generate new configurations efficiently\cite{baek2021accurate, mirdita2022colabfold,jumper2021highly, evans2021protein, rives2021biological, lin2023evolutionary,lewis2025scalable,abramson2024accurate,watson2023novo, lu2024str2str, morehead2024geometry,chu2024all}. Score-based diffusion models (SBDM) are particularly attractive because they define a flexible sampling process that can represent complex molecular ensembles\cite{hoogeboom2022equivariant,sohl2015deep,ho2020denoising,song2021score,watson2023novo,lu2024str2str,morehead2024geometry,chu2024all}. 
However, most transferable molecular generative models are trained to reproduce the distribution of a single molecule\cite{watson2023novo,lu2024str2str,lewis2025scalable,hoogeboom2022equivariant,chu2024all}. Some generative models aim to sample molecular structures conditional on a fixed external environment, such as protein pockets \cite{guan2023targetdiff,peng2022pocket2mol,ragoza2022generating} and material surfaces \cite{kolluru2024adsorbdiff,ronne2024generative}. Most generative models trained for multicomponent systems focus on protein-protein complexes \cite{watson2023novo,ketata2023diffdockpp,song2025ppdiff}. Similarly, only a few generative models are capable of assembling multiple fragments for specific applications. A general theory for sampling structures of a multicomponent system has not yet been developed. 

When multiple components or fragments are generated independently, the product of their learned distributions does not automatically represent the physically correct interacting ensemble. The missing ingredient is the interaction energy that couples the separately learned molecular distributions.
Consider two molecular units that interact through covalent and/or non-covalent interactions, such as two molecules in a binding complex or two fragments in a macromolecule. The Cartesian coordinates of these two units are denoted by $\ux$ and $\uy$. 
If two independently trained generative models represent molecular unit distributions $P_1(\ux)$ and $P_2(\uy)$, independent sampling gives the uncoupled distribution $P_1(\ux)P_2(\uy)$. The physically relevant coupled ensemble, however, should include an interaction potential $\uint(\ux,\uy)$, such that the target distribution is proportional to
\begin{equation}
    P(\ux,\uy)\propto P_1(\ux)P_2(\uy)\exp[-\beta \uint(\ux,\uy)]\;\ldotp
    \label{eq:int}
\end{equation}
where $\beta=1/k_{\rm B}T$ and $T$ is the temperature. 
This expression suggests a modular strategy: learned generative models provide structural priors for individual components, while explicit physical interactions define how these priors should be combined. Such a strategy would avoid retraining a new generative model for every coupled multicomponent system and would instead allow separately trained models to be reused and assembled through physics-based coupling\cite{wang2025extrapolating}. 

Here, we introduce physical-interaction-coupled diffusion models (PICDiff), a framework for composing independently trained SBDMs through bonded and nonbonded interaction potentials during reverse diffusion. PICDiff leaves the pretrained SBDMs unchanged, while additional interaction forces guide the reverse diffusion process toward configurations consistent with the coupled molecular Hamiltonian. This approach provides a physics-guided route for sampling conformations of a multicomponent system from separately learned samplers for all components. 

We demonstrate PICDiff in representative coupling problems involving both nonbonded and bonded interactions. In the first case, separately trained peptide and polymer SBDMs are coupled through nonbonded interactions to sample a peptide-polymer interface. In the second case, short polymer-fragment SBDMs are coupled through bonded and nonbonded interactions to construct longer-chain polymer ensembles. These examples test whether physical interaction guidance can recover thermodynamic observables, conformational rearrangements, and structural statistics of the full coupled systems. The manuscript is organized as follows. We first briefly review the theory of SBDM and then introduce the theoretical framework of PICDiff. We then present two examples: sampling the conformations of a peptide–polymer binding complex and sampling the conformations of a long polymer chain by decomposing it into fragments.

\section{METHODS}
\label{sec:method}
\subsection{A Brief Introduction to Score-Based Diffusion Models}
Here, we briefly summarize the theory of SBDM. During SBDM training, noise is added to a protein structure through the forward SDE
\begin{equation}
    \rd \mathbf{x}=\uf(\mathbf{x}, t)\, \rd t+g(t)\, \rd \mathbf{w}, \qquad t\in [0,1],
    \label{eq:foward_sde}
\end{equation}
where $\uf(\mathbf{x}, t)$ and $g(t)$ are the drift and diffusion terms, and $\mathbf{w}$ is a Wiener process. In this work, we use $\uf(\mathbf{x}, t)=0$ and $g(t)=\sqrt{\mathrm{d}\sigma^2(t)/\mathrm{d}t}$, where $\sigma(t)$ is the noise scale. The corresponding reverse-time SDE is \cite{anderson1982reverse,song2020score}
\begin{equation}
    \rd \mathbf{x} = [\uf(\mathbf{x},t) - g^{2}(t)\nabla_{\mathbf{x}}\log P(\mathbf{x},t)]\rd t +g(t)\rd\bar{\mathbf{w}},
    \label{eq:rev_SDE}
\end{equation}
where $\bar{\mathbf{w}}$ is a reverse-time Wiener process. The score function $\nabla_{\mathbf{x}}\log P(\mathbf{x},t)$ is approximated by a neural network $\mathbf{s}_{\boldsymbol{\theta}}(\mathbf{x}(t), t)$ parametrized by $\boldsymbol{\theta}$. To train the score function, we can use a weighted dataset $\{\tilde{w}_i,\ux(0)^{(i)}\}$ where $\ux(0)^{(i)}$ is the $i$th configuration with weight $\tilde{w}_i$. $\tilde{w}_i=1$ for an unweighted dataset. 
The weighted loss is then defined as
\begin{widetext}
\begin{equation}
\mathcal{L}_{\mathrm{weight}}(\theta)
=
\sum_{i=1}^{B} \tilde{w}_i \,
\mathbb{E}_{\ux(t) \sim q(\ux(t) \mid \ux(0)^{(i)})}
\left[
\sigma^2(t)
\left\|
s_{\theta}(\ux(t),t)
-
\nabla_{\ux(t)} \log q_t\!\left(\ux(t) \mid \ux(0)^{(i)}\right)
\right\|^2
\right]\;\ldotp
\label{eq:weighted_loss}
\end{equation}
\end{widetext}

\subsection{Coupling Diffusion Models with Physics}
\label{sec:picdiff}

To simplify the discussion, we assume that the molecular system is composed of two units, although generalization to systems with multiple units is straightforward. For two non-interacting units, two baseline SBDMs,
\begin{subequations}
\begin{align}
    \rd \ux & = [\uf_1(\ux,t) - g^{2}(t)\us_1(\ux,t)]\rd t +g(t)\rd\bar{\mathbf{w}}_1, \\
    \rd \uy & = [\uf_2(\uy,t) - g^{2}(t)\us_2(\uy,t)]\rd t +g(t)\rd \bar{\mathbf{w}}_2, \
\end{align}
\end{subequations}
sample approximate Boltzmann distributions of two units: $P_1(\ux) \propto \exp[-\beta U_1(\ux)]$ and $P_2(\uy) \propto \exp[-\beta U_2(\uy)]$, where $U_1(\ux)$ and $U_2(\uy)$ are effective internal potentials for two units. If we require the two SBDMs to share the same $g(t)$, we can define a joint diffusion model 
\begin{equation}
    \mathrm{d} \uX = [\uF(\uX,t) - g^{2}(t)\uS(\uX,t)]\mathrm{d}t+g(t)d\bar{\mathbf{W}}
    \label{eq:diff-joint}    
\end{equation}
where $\uX=(\ux,\uy)$ denotes the coordinates of the complex of two units, respectively. $\uS=(\us_1,\us_2)$, $\uF=(\uf_1,\uf_2)$, and $\bar{\mathbf{W}}=(\bar{\mathbf{w}}_1,\bar{\mathbf{w}}_2)$ denote the corresponding score functions, time-dependent drift terms, and reverse-time Brownian motions.

\begin{figure}[ht]
    \centering
    \includegraphics[width=\columnwidth]{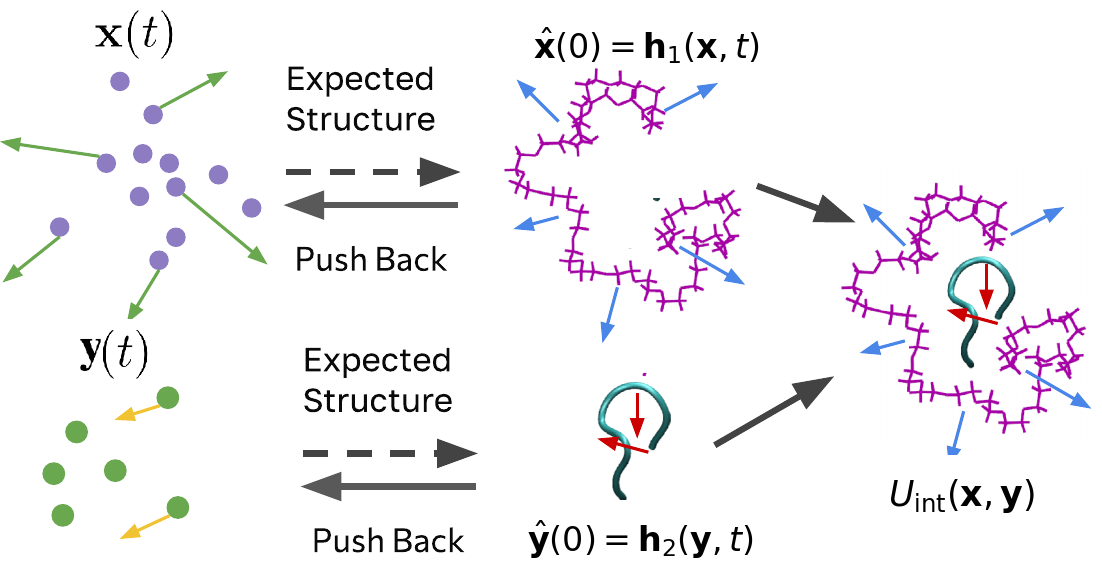}
    \caption{Schematic of PICDiff inference for peptide--polymer interface through non-bonded interactions. Expected structures $\hat{\ux}(0)$ and $\hat{\uy}(0)$ are estimated from noisy coordinates $\ux(t)$ and $\uy(t)$. A coupling potential $\uint(\hat{\ux}(0),\hat{\uy}(0))$ is evaluated. Blue and red arrows indicate coupling forces on the polymer and protein, respectively. These forces are mapped back to time-$t$ coordinates as drift terms, shown as green (polymer) and yellow (peptide) arrows.}
    \label{fig:pgdg-f}
\end{figure}

In PICDiff, coupling is introduced through an additional term $\uint(\ux,\uy)$, so that the distribution of the interacting system becomes $p^{\mathrm{PIC}}_{\mathrm{B}}(\ux,\uy) \propto \exp[-\beta (U_1(\ux)+U_2(\uy)+\uint(\ux,\uy))]$. 
To sample $p^{\mathrm{PIC}}_{\mathrm{B}}$, we can modify Eq.(\ref{eq:diff-joint}) by adding an extra drift term \cite{wang2403protein}:
\begin{widetext}
\begin{equation}
    \rd \mathbf{\uX} = \left[\uF(\uX,t) - g^{2}(t)\left(\uS(\uX,t)
    - \nabla_{\uX(t)}U_t(\uX(t),t)\right)\right]\rd t +g(t)d\bar{\mathbf{W}},
    \label{eq:rev_SDE_dlh}
\end{equation}
\end{widetext}
where 
\begin{equation}
    U_t(\mathbf{X}(t))=-k_{\mathrm{B}}T\log
    \mathbb{E}_{p(\uX(0)|\uX(t))}
    \left(e^{-\beta \uint(\uX(0))}\right)\;\ldotp
    \label{eq:ut}
\end{equation}
The challenge of using Eq.(\ref{eq:ut}) is the lack of a closed-form expression for evaluating $\mathbb{E}_{p(\uX(0)|\uX(t))}\left(\cdot\right)$. Following the manifold-constraint method \cite{chung2022improving,chung2022diffusion}, we approximate $p(\uX(0)|\uX(t))\approx \delta(\uX(0)-\hat{\uX}(0))$, where $\hat{\uX}(0)=\mathbb{E}[\uX(0)|\uX(t)]$. For DDPM-like models, $\hat{\uX}(0)$ can be written explicitly as $\hat{\uX}(0)=\mathbf{h}(\uX(t),t)$ \cite{ho2020denoising}, which gives $U_t(\uX(t))\approx \uint(\mathbf{h}(\uX(t),t))$. The reverse SDE in PICDiff is therefore written as
\begin{widetext}
\begin{equation}
    \rd \uX = \left[\uF(\uX,t) - g^{2}(t)\left(\uS(\uX,t)
    - \lambda(t)\nabla_{\uX(t)}\uint(\mathbf{h}(\uX(t),t))\right)\right]\rd t +g(t)\rd\bar{\mathbf{W}},
    \label{eq:rev_SDE_dlh}
\end{equation}
\end{widetext}
where $\lambda(t)$ controls the coupling strength. 
Since the approximation above becomes more accurate at small $t$, $\lambda(t)$ is chosen to be weak near $t=1$ and stronger as $t\to 0$. In practice, $\hat{\uX}(0)=(\hat{\ux}(0),\hat{\uy}(0))=(\mathbf{h}_1(\ux(t),t),\mathbf{h}_2(\uy(t),t))$ where $\hat{\ux}(0)$ and $\hat{\uy}(0)$ are expected structures from the two SBDMs (see SI for details of $\uh_1$ and $\uh_2$). The schematic of PICDiff is presented in Fig.(\ref{fig:pgdg-f}). 
In this way, PICDiff biases the sampling toward structures adapted to the coupling interactions without retraining the foundation model. 

The effective energy function corresponding to $P(\ux,\uy)$ can be written as $U(\ux,\uy)=-\kb T P(\ux,\uy)$ and $\uint(\ux,\uy)=U(\ux,\uy)-U_1(\ux)-U_2(\uy)$. In practice, $\uint(\ux,\uy)$ can be modeled with various approximations. For example, residue-level peptide-polymer interactions are modeled by model systems of dipeptides binding with short polymer chains. In the second example, $\uint(\ux,\uy)$ that connects polymer fragments takes the functional form of bounded and nonbounded interactions from a conventional force field \cite{charmm, amber, OPLSAA}.  

\subsection{Unit Alignment for Stiff Interactions}
\label{sec:align}

Typically, score functions in SBDM are equivariant, so there is no control on the position and orientation of each individual unit. In PICDiff, $\uint$ can determine the energetically preferred positions and orientations. However, if $\uint$ is stiff and if two units are misaligned, the forces of $\uint$ will be large, and it may cause the numerical integration of Eq.(\ref{eq:rev_SDE_dlh}) to be unstable. Specifically, this could be an issue if $\uint$ involves harmonic bonded interactions. 
In this section, we extend the method to efficiently couple multiple units through bonded interactions. The key development is to introduce a relative alignment among different diffusion-model-generated units to minimize the forces from $\uint$ and stabilize PICDiff inference. 

\begin{figure}[ht]
    \centering
    \includegraphics[width=\columnwidth]{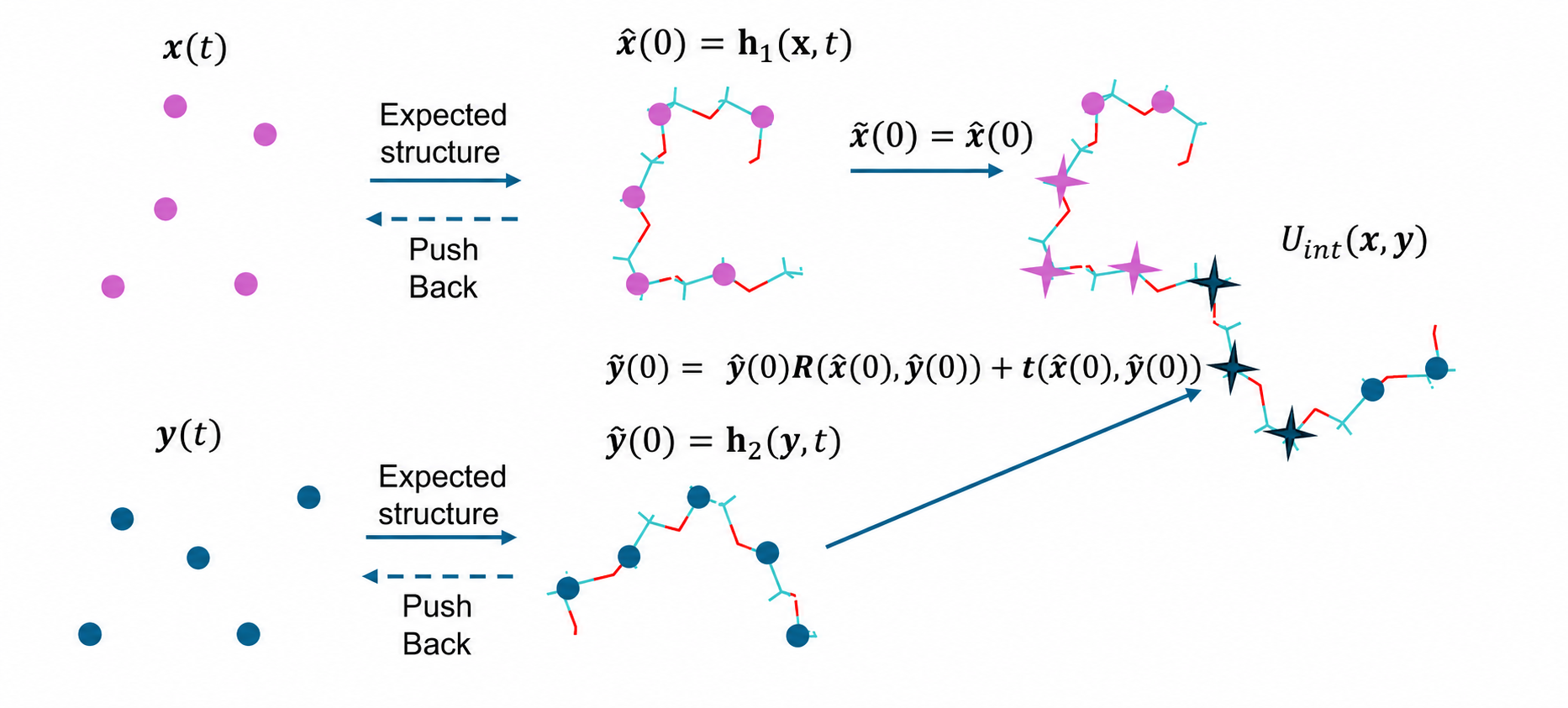}
    \caption{Schematic of PICDiff inference for polymer chains with bonded and non-bonded interactions. The expected structures $\hat{\ux}(0)$ and $\hat{\uy}(0)$ are predicted from noisy inputs $\ux(t)$ and $\uy(t)$. A coupling potential $\uint(\hat{\ux}(0),\hat{\uy}(0))$ is evaluated, and the relative posture is determined by applying rotational and translational refinement to $\hat{\uy}(0)$, producing $\tilde{\uy}(0)$, so as to minimize the bonded and nonbonded coupling potential. The residual interaction after posture refinement is taken as the coupling force, shown as blue arrows on the first polymer block and red arrows on the second polymer block.
    These forces are mapped back to the time-$t$ coordinates as drift terms, shown as green arrows for $\ux(t)$ and yellow arrows for $\uy(t)$.}
    \label{fig:bonded-coupling}
\end{figure}

Similar to Sec. (\ref{sec:picdiff}), we will demonstrate the idea of unit alignment with a system made of two units (Fig.(\ref{fig:bonded-coupling})). Consider a system composed of two molecular units with coordinates $\mathbf{X}=(\ux,\uy)$. 
Following the discussion in Sec.(\ref{sec:picdiff}), at diffusion time $t$, the generative model predicts the corresponding denoised structure $(\hat{\ux}_0,\hat{\uy}_0)=(\mathbf{h}_1(\ux(t),t),\mathbf{h}_2(\uy(t),t))$ using Tweedie's formula. We treat $\hat{\ux}(0)$ and $\hat{\uy}(0)$ as rigid bodies and optimize their relative poses to minimize $\uint$. 
Specifically, $\hat{\ux}(0)$  is kept fixed while $\hat{\uy}(0)$ is transformed by a rigid-body motion
\begin{equation}
\tilde{\uy}(0)=\hat{\uy}(0)\mathbf{R}^{\top}+\mathbf{t},
\end{equation}
where $\mathbf{R}$ is a rotation matrix and $\mathbf{t}$ is a translation vector. 
Optimal $\mathbf{R}$ and $\mathbf{t}$ are obtained by minimizing $\uint(\tilde{\ux}(0)\equiv\hat{\ux}(0),\tilde{\uy}(0))$. With optimized $\mathbf{R}^\ast$ and $\mathbf{t}^\ast$, the guidance term in Eq.(\ref{eq:rev_SDE_dlh}) becomes 
\begin{equation}
    \nabla_{\uX(t)}\uint(\mathbf{h}_1(\ux(t),t),\mathbf{h}_2(\uy(t),t)\mathbf{R}^{\ast\top}+\mathbf{t}^\ast)\;\;\ldotp
\end{equation}
We emphasize that minimizing $\uint$ is used only to stabilize the inference of SBDM. Therefore, only loose convergence criteria are needed, and the computational overhead introduced by the minimization is limited. 

\section{RESULTS}
\subsection{$A\beta_{\mathrm{16-22}}$ at PEG interface} 
We first test PICDiff using physical coupling that includes only nonbonded interactions, focusing on the interaction between the amyloid-\(\beta\) fragment A\(\beta_{16-22}\) (sequence KLVFFAE) and polyethylene glycol (PEG) with 28 monomers (PEG$_{28}$). 
A\(\beta_{16-22}\) is a commonly studied amyloid-forming peptide fragment and has been widely used as a model system for peptide aggregation and interfacial molecular simulations \cite{wang2019thermodynamic, okumura2020molecular}. 
PEG is among the most widely used synthetic polymers in biomedical and molecular systems due to its water solubility and generally favorable biocompatibility \cite{knop2010polyethylene,harris2003effect,padin2022understanding}. 
Thus, the peptide-PEG system provides a representative model for protein-polymer interfaces, where PEG can modulate protein adsorption, protein-protein interactions, and biomolecular interfacial behavior \cite{michel2005influence,kulkarni2000effects,wu2013investigation,perera2021understanding}.
Unlike large proteins with well-folded structures, PEG binding to $A\beta_{\mathrm{16-22}}$ is a challenging example for PICDiff since both peptide and PEG can have significant conformational changes during the binding process. 
$A\beta_{\mathrm{16-22}}$ contains hydrophobic residues (Leu17, Val18, Phe19, and Phe20), as well as charged residues (Lys16, Glu22). 
As a result, adsorption between $A\beta_{\mathrm{16-22}}$ and PEG is governed by multiple competing factors, including hydrophobic interactions, electrostatic effects, and chain conformational flexibility. This makes the adsorption behavior difficult to predict empirically.

\subsubsection{Training SBDMs for Peptide and Polymer}
\label{sec:pep-peg-sbdm1}
We first trained an SBDM, named $\mathrm{DM}_{\mathrm{PRO}}$, for \(A\beta_{16-22}\).  
Because $A\beta_{16-22}$ may undergo substantial conformational changes upon binding to PEG$_{28}$, $\mathrm{DM}_{\mathrm{PRO}}$ must be able to sample a broad range of peptide conformations. Training $\mathrm{DM}_{\mathrm{PRO}}$ therefore requires a dataset containing diverse $A\beta_{16-22}$ structures. However, extended conformations with highly exposed hydrophobic residues are only weakly populated in plain MD simulations. To enhance conformational diversity, we augmented the peptide training data with enhanced-sampling simulations. Specifically, well-tempered metadynamics (WTM) \cite{barducci2008well} was performed using stochastic kinetic embedding, an ML-based collective variable \cite{zhang2018unfolding}, to promote peptide conformational exploration (see Supporting Information for details of the collective-variable construction and WTM simulation details). Although this procedure increased conformational diversity, the sampled ensemble remained dominated by compact structures with relatively small radii of gyration ($R_g$). Therefore, an additional trajectory enriched in large-$R_g$ conformations was included in the training dataset (see Supporting Information for details of this simulation).
We used the architecture of the DiffAb \cite{luo2022antigen} model for $\mathrm{DM}_{\mathrm{PRO}}$. In $\mathrm{DM}_{\mathrm{PRO}}$, protein structures are represented in residue-specific local frames, and the diffusion process is defined over residue-wise translations and rotations. 
The reweighting factor for a structure generated by $\mathrm{DM}_{\mathrm{PRO}}$ was calculated as 
\begin{equation}
    w\propto p_{\mathrm{eq}}/p_{\mathrm{data}}
    \label{eq:reweight}
\end{equation}
where $p_{\mathrm{eq}}$ is the equilibrium probability and $p_{\mathrm{data}}$ is the probability of the training data. Both $p_{\mathrm{eq}}$ and $p_{\mathrm{data}}$ were evaluated using kernel density estimation. For each sampled structure, $p_{\mathrm{data}}$ was calculated by comparing the structure with all configurations in the training dataset, whereas $p_{\mathrm{eq}}$ was calculated by comparing it with the configurations obtained from the reweighted WTM ensemble.


For PEG$_{28}$, sampling is achieved by a WTM simulation with $R_g$ as the collective variable. However, these configurations are used as training data without assigning weights. We used the biased ensemble for diffusion model training to avoid suppressing low-probability extended conformations. 
The diffusion model trained with this dataset is denoted as $\mathrm{DM}_{\mathrm{PEG}}$ (see Supporting Information for details of the WTM simulation for PEG$_{28}$ and details of training $\mathrm{DM}_{\mathrm{PEG}}$). 
We also used the DiffAb architecture for $\mathrm{DM}_{\mathrm{PEG}}$, treating each -O-C$^\alpha$H$_2$-C$^\beta$H$_2$- unit as a ``residue''. 
The translational center of each residue is defined by the position of C$^\alpha$. 
To unbias the ensemble of generated structures, each generated PEG structure is assigned a weight using Eq.(\ref{eq:reweight}) with $p_{\rm eq}$ and $p_{\rm data}$ calculated from kernel density estimation. 

\subsubsection{Developing a Peptide-PEG Interaction Potential}
After training $\mathrm{DM}_{\mathrm{PRO}}$ and $\mathrm{DM}_{\mathrm{PEG}}$ independently, 
physical interactions $U_{\mathrm{int}}$ are introduced to couple the two models during inference. 
As in our previous work, intermolecular guidance is applied only to translational sites, not to residue orientations. This approximation avoids unstable guidance on $SO(3)$, where the standard Tweedie-based denoising estimate is not directly applicable \cite{chung2022diffusion}.

Conventionally, CG interactions between PEG and protein are modeled using pairwise interactions \cite{ramezanghorbani2018optimizing,lee2009coarse}. However, pairwise interaction may not capture the non-additive changes of solvation free energy during the binding process of PEG and protein \cite{sanyal2016coarse,john2017many}. Therefore, we designed a residue-level many-body CG interaction potential. 
We parameterize the residue-level CG interaction with MD simulations of a dipeptide molecule binding to a short PEG chain containing five monomer units (PEG$_{5}$). The total peptide-PEG interaction is then approximated as the sum of residue-based interactions. 
As illustrated in Fig.~\ref{fig:fig:peo-interaction}(a), each dipeptide-PEG$_{5}$ reference system is projected onto a set of 9 pairwise distances between three carbon atoms ($\alpha$-carbon and two methyl carbons from the ACE and NME caps) and 3 sites on PEG PEG$_{5}$. 
The capped CH$_3$ groups are used to mimic the neighboring residue C$_\alpha$ sites in the peptide, which is similar to the strategy used in our previous work \cite{wang2025extrapolating}. In this way, the many-body residue-polymer interactions also include the effect of the relative orientation between a residue and a short PEG fragment. 

The many-body dipeptide–PEG$_5$ interaction is a statistical potential with the corresponding probability distribution modeled by a Gaussian mixture model (GMM). Because GMM-based probability estimates are less reliable for low-probability conformations, particularly in regions dominated by short-range excluded-volume effects, we further introduce a short-range pairwise repulsive potential to prevent steric clashes during generated-structure sampling. When applying the dipeptide–PEG$_5$ interaction, the positions of carbon atoms in the ACE and NME caps are replaced by the $\alpha$-carbon positions in neighboring residues. 
$U_{\mathrm{int}}$ is the sum of the dipeptide–PEG$_5$ interactions between all peptide residues and all PEG$_5$ fragments. 
Details of this modeling procedure are provided in the Supporting Information.

\subsubsection{Calculation of the Binding Potential of Mean Force (PMF)}
\begin{figure}[ht]
    \centering
    \includegraphics[width=\columnwidth]{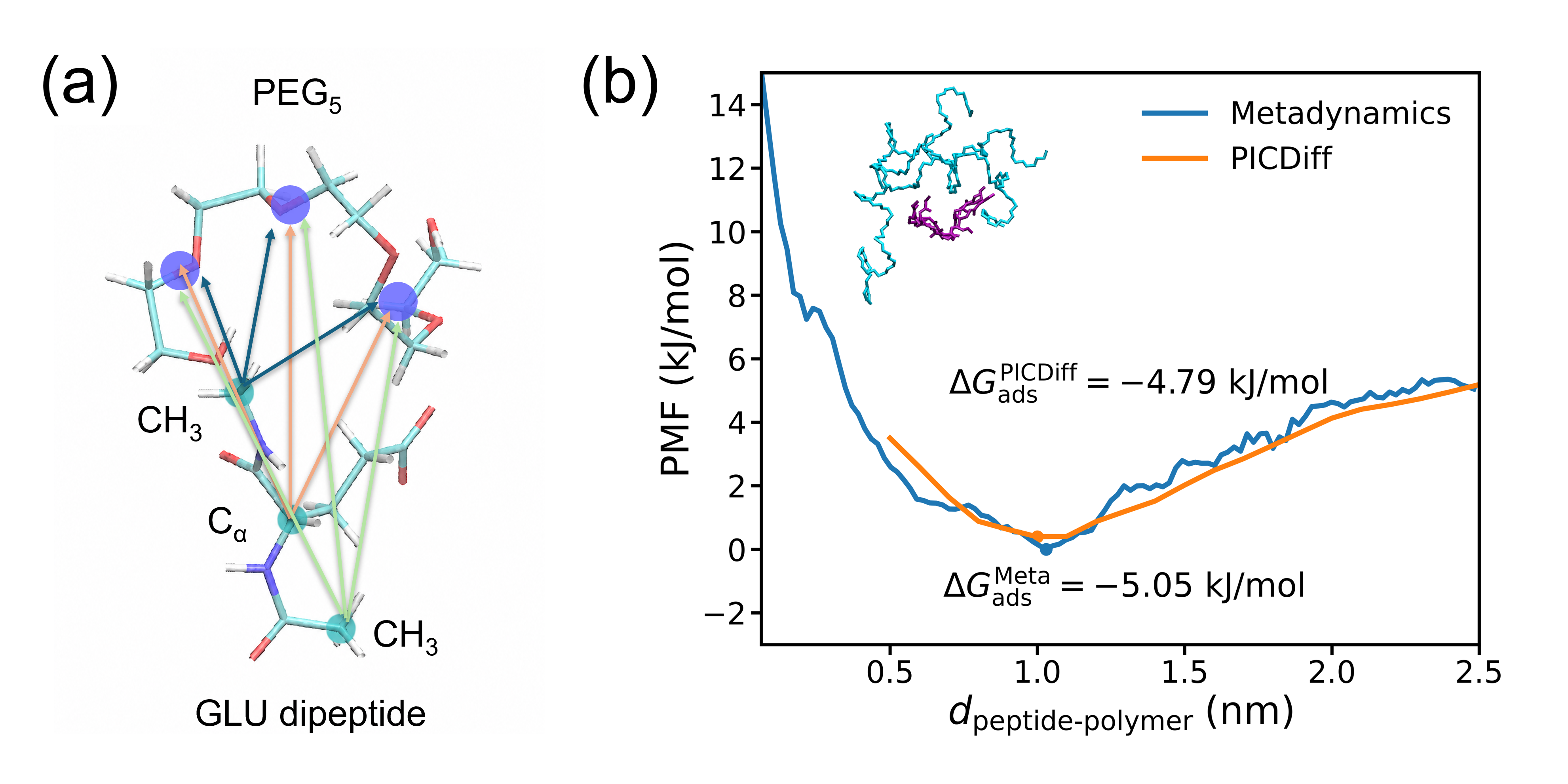}
    \caption{
    {\bf (a)} A residue–PEG interaction potential is modeled by an MD simulation of a dipeptide binding with PEG$_5$. The potential is a function of all 9 distances presented as arrowheads in the insert, formed between three dipeptide sites, two CH$_3$  and one C$\alpha$ atom, and three PEG$_5$ sites, defined as the geometric centers of C$^\alpha$ 1--3, 2--4, and 3--5.
    {\bf (b)} Adsorption free energy profile along the peptide–PEG separation distance. Comparison between enhanced sampling (blue) and PICDiff (orange) shows that PICDiff accurately reproduces the binding thermodynamics. The inset shows a representative interfacial ensemble.}
    \label{fig:peo-interaction}
\end{figure}

$A\beta_{\mathrm{16-22}}$-PEG$_{28}$ configurations sampled by PICDiff were reweighted with the weights generated in Sec.(\ref{sec:pep-peg-sbdm1}) (See Supporting Information for details of the weight calculation). 
One advantage of PICDiff is that the binding PMF can be calculated directly from $U_{\mathrm{int}}$ and the ensemble generated by PICDiff. 
Using $A\beta_{\mathrm{16-22}}$-PEG$_{28}$ as an example, we computed the PMF along the center-of-mass distance between $A\beta_{\mathrm{16-22}}$ and PEG$_{28}$ (\(d_{\mathrm{peptide-polymer}}\)). For PICDiff, the PMF was obtained using thermodynamic integration (TI) \cite{frenkel2002understanding,klimovich2015guidelines}, in which the free-energy difference is reconstructed by integrating the mean force along the reaction coordinate,
\begin{equation}
A(d) = A(d_0) + \int_{d_0}^{d_{\mathrm{peptide-polymer}}}
\left\langle
\frac{\partial U_{\mathrm{int}}}{\partial d'}
\right\rangle_{d'}
\, \mathrm{d}d' \;\;\ldotp
\end{equation}
Here, \(d_0\) is the smallest $d_{\mathrm{peptide-polymer}}$. 
The average \(\langle \cdots \rangle_{d'}\) is evaluated over the ensemble generated by PICDiff at fixed \(d'\). We want to emphasize that optimal binding orientations of $A\beta_{\mathrm{16-22}}$ and PEG$_{28}$ are automatically determined by PICDiff. 
The PMF predicted by PICDiff agrees well with that obtained from the WTM simulation (details of the validation procedure are provided in the Supporting Information). In particular, PICDiff correctly predicts that the most stable binding distance is approximately \(1~\mathrm{nm}\) and predicts the binding free energy with an error of less than \(1~\mathrm{kJ\,mol^{-1}}\)(See Fig.~\ref{fig:peo-interaction}(b)). 

\subsubsection{Evaluating PICDiff-Sampled Conformations}
\begin{figure*}[ht]
    \centering
    \includegraphics[width=\textwidth]{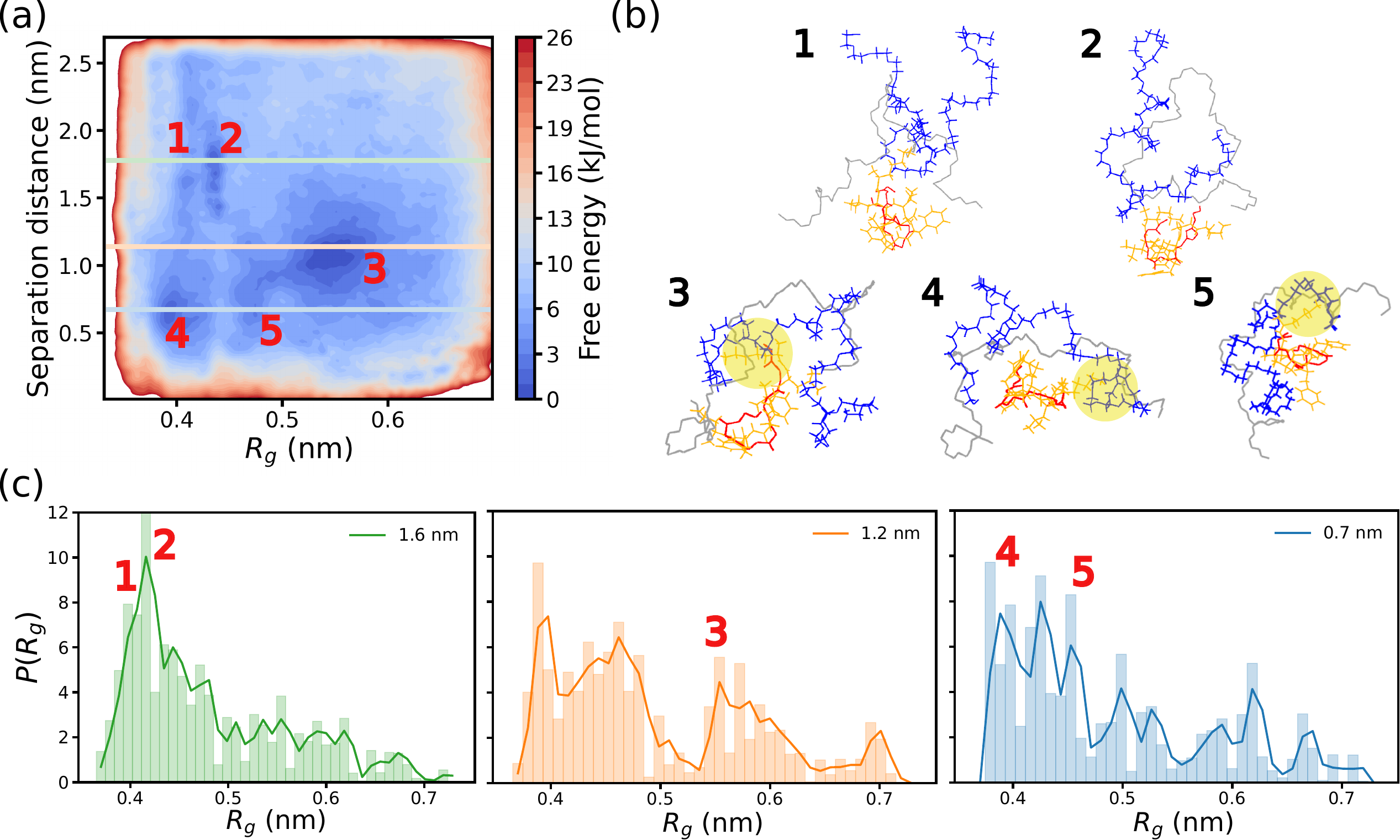}
    \caption{{\bf(a)} $A\beta_\mathrm{16-22}$ peptide at the PEG$_{28}$ interface obtained from WTM simulations, using the $R_g$ of $A\beta_\mathrm{16-22}$ and the center-of-mass distance ($d_{\mathrm{peptide-polymer}}$) as collective variables. Five metastable conformations are identified and labeled. 
    {\bf(b)} Representative structures of the five metastable conformations. Structures sampled by PICDiff and WTM are overlaid. For $A\beta_\mathrm{16-22}$, structures generated by PICDiff are shown in red, while those from WTM are shown in orange. For PEG$_{28}$, PICDiff-sampled structures are shown in grey, and WTM-sampled conformations are shown in blue. For structures 3–5, the interaction region between the Lys residue and the PEG chain is highlighted in yellow. 
    {\bf(c)} $R_g$ distributions of $A\beta_\mathrm{16-22}$ peptide in the complexes sampled by PICDiff at fixed $d_{\mathrm{peptide-polymer}}$ of 0.7 nm, 1.2 nm, and 1.6 nm. Peaks corresponding to the five metastable conformations are identified and labeled at each distance.}
    \label{fig:structures}
\end{figure*}

Because quantitative calculation of the binding PMF requires accurate sampling of the ensemble distribution, we also evaluate the quality of ensemble sampling. 
We first analyzed the WTM simulations by projecting the sampled trajectories onto a two-dimensional CV space spanned by the peptide \(R_g\) and $d_{\mathrm{peptide-polymer}}$, 
as shown in Fig.~\ref{fig:structures}(a). 
We identified 5 metastable conformations on the free energy surface, which represent the major conformational changes of \(A\beta_{\mathrm{16-22}}\) upon binding to PEG$_{28}$. 
Representative $A\beta_{\mathrm{16-22}}$-PEG$_{28}$ structures from the WTM simulation and from PICDiff were selected for each metastable conformation (Fig.~\ref{fig:structures}(b)).
The PICDiff-generated and WTM-sampled structures were aligned by minimizing the peptide RMSD.

For conformations 1 and 2, $d_{\mathrm{peptide-polymer}}$ is relatively large. Contacts between the peptide and polymer are limited and localized, while the non-contact regions of PEG$_{28}$ remain highly flexible. Even in these cases, PICDiff is able to predict the residue-level contact regions accurately.
For conformations 3 to 5, $d_{\mathrm{peptide-polymer}}$ becomes smaller, allowing extensive interactions between the peptide and polymer. PICDiff captures key features of the binding mechanism. As shown in the representative structures of conformations 3, 4, and 5, the highlighted yellow regions in Fig.~\ref{fig:structures}(b) indicate the key binding sites identified from the WTM simulations. In particular, the positively charged $\mathrm{NH_3^+}$ at Lys sidechain and the N-terminus form strong electrostatic interactions with PEG oxygen atoms. The PEG chain tends to encircle the $\mathrm{NH_3^+}$ groups. Notably, PICDiff uses a model without explicit sidechain resolution, so the emergence of these binding modes is driven entirely by $\uint$. 
As shown in Fig.~\ref{fig:structures}(c), three representative values of $d_{\mathrm{peptide-polymer}}$, $1.6$, $1.2$, and $0.7~\mathrm{nm}$, were selected to characterize the peptide $R_g$ distributions. At $d_{\mathrm{peptide-polymer}}\approx1.6~\mathrm{nm}$, corresponding to conformations 1 and 2, the distribution shows two high-probability peaks near $R_g \approx 0.4~\mathrm{nm}$. At $d_{\mathrm{peptide-polymer}}\approx1.2~\mathrm{nm}$, mainly corresponding to conformation 3, the probability density near $R_g \approx 0.55~\mathrm{nm}$ becomes more pronounced. At $d_{\mathrm{peptide-polymer}}\approx0.7~\mathrm{nm}$, corresponding to conformations 4 and 5, the distribution shifts toward more compact conformations, with preferred $R_g$ values between $0.4~\mathrm{nm}$ and $0.5~\mathrm{nm}$. 

\subsection{Sampling Polymer Conformations with PICDiff}
In the previous section, we demonstrated the capability of PICDiff to sample structures of peptide-polymer complex using non-bonded physical interactions. Here, we extend PICDiff to include both bonded interactions and nonbonded interactions to couple fragments of a polymer molecule. 
With this design, PICDiff can sample the conformations of a long polymer chain, which is crucial for developing functional polymers such as biomimetic polymers \cite{turney2025atomistic,lutz2013sequence,torkelson2024rational,sun2013peptoid,ganewatta2021chemical,webb2020targeted} and sequence-to-morphology prediction\cite{bishnu2025morphology,park2023prediction}. 
The challenge of sampling polymer conformations arises from the high energy barriers associated with conformational equilibration. 
Recently, efficient generation of polymer configurations has attracted growing attention as a means of overcoming this bottleneck. Existing efforts have pursued different strategies, including developing hierarchical generative architectures tailored to polymer conformations \cite{wang2025polyconf,simm2025simpoly}. 
Training a transferable model to sample polymer conformations is also challenging because the chemical space required to cover diverse polymer and copolymers units is large, making it difficult to prepare a sufficiently broad dataset \cite{wu2020polymerinformatics,cencer2022polymerml,kim2023openmacromoleculargenome}. 
PICDiff is designed to solve this problem by coupling short polymer fragments represented by SBDMs to sample conformations of a long chain. 
We use PEG chains with different lengths as a demonstration. Although PEG is relatively simple, modeling its conformations remains important in biomedical applications \cite{knop2010polyethylene, harris2003effect, padin2022understanding}. 
While we focus here on PEG as a demonstration, the PICDiff framework can, in principle, be generalized to more complex systems, such as copolymers. 

\subsubsection{Developing SBDM and Inter-fragment Interactions}
In this example, we tested generating PEG$_{10}$ from two PEG$_{5}$ fragments and generating PEG$_{20}$ from two PEG$_{10}$ fragments. 
To train a SBDM for PEG$_{5}$ fragments, a WTM simulation of PEG$_{5}$ in aqueous solution with $R_g$ as CV was performed to collect the training data. 
For the PEG$_{10}$ SBDM, we adopted a different strategy for training-data generation. Rather than using physics-based simulations, we directly used PEG$_{10}$ structures sampled by PICDiff. This choice was motivated by our goal of evaluating whether PICDiff can enable hierarchical modeling, where PICDiff is applied iteratively to generate PEG$_{2^m n}$ from PEG$_n$. 
The same SBDM design as in Sec.(\ref{sec:method}) was used. 
Details of SBDM training are provided in the Supporting Information. 
\(U_{\mathrm{int}}\), including both bonded and nonbonded terms, was employed to couple two separate SBDMs.
Specifically, the bonded part of \(U_{\mathrm{int}}\) is defined on C$^\alpha$ near the junction between two adjacent fragment models (Fig.(\ref{fig:bonded-coupling})). 
Assuming an interface between two fragments occurs between unit $i$ and unit $i+1$, the bonded interactions contain all bond, bend, and torsion interactions that include C$^\alpha_i$ and C$^\alpha_{i+1}$. 
The bond, bend, and torsion energy terms are fitted from the configurational ensemble of PEG$_5$ sampled by metadynamics simulations. We use the functional forms of bonded interactions in conventional force fields \cite{charmm,amber,OPLSAA}, such as harmonic bond energy and harmonic bend energy. 
The nonbonded interactions in \(U_{\mathrm{int}}\) assign short-range repulsive interactions between monomer translational sites to prevent unphysical overlap among fragments. 
Details of the junction definition and potential parameterization are provided in Supporting Information. 
Because the bounded interactions contain stiff harmonic interactions, the unit alignment method introduced in Sec.(\ref{sec:align}) was applied. 

\subsubsection{Sampling Polymer Structures with PICDiff}
\begin{figure*}[ht]
    \centering
    \includegraphics[width=\textwidth]{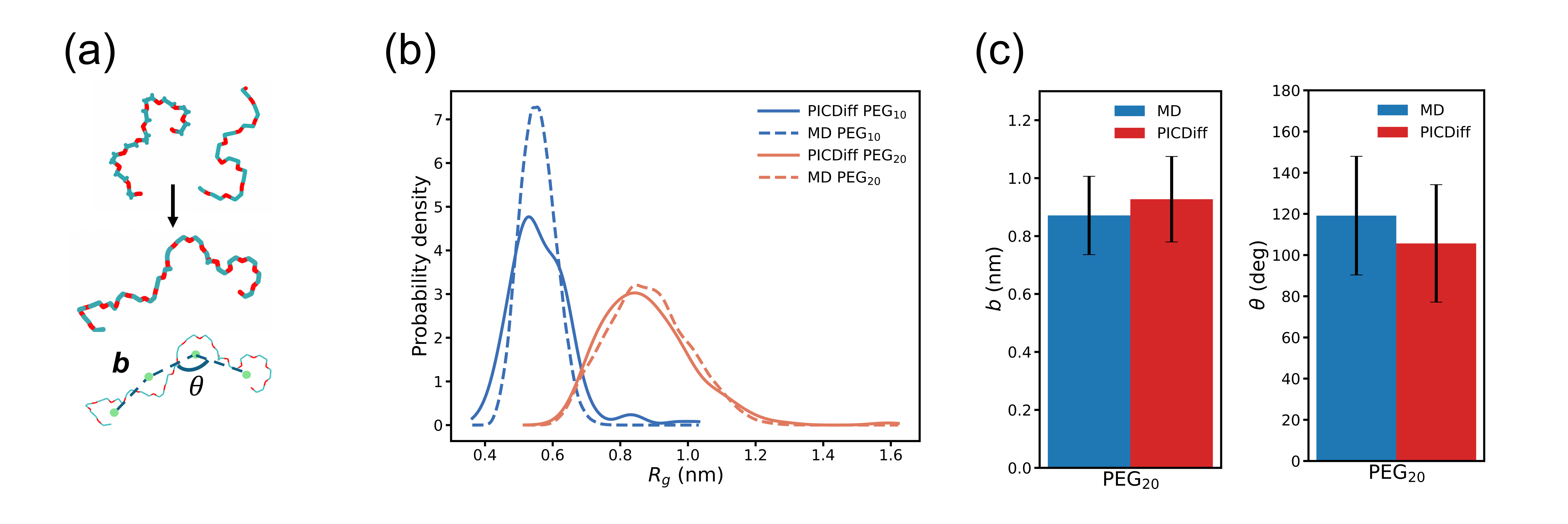}
    \caption{{\bf(a)} Schematic illustration of long PEG chain generation from short fragments. The inter-fragment distance between neighboring fragments, denoted as $b$, and the inter-fragment angle defined by three consecutive fragments, denoted as $\theta$, are used to characterize the PEG chain conformations. {\bf(b)} Radius of gyration ($R_g$) distributions of PEG$_{10}$ and PEG$_{20}$ obtained from PICDiff-generated samples and WTM simulations. {\bf(c)} Averaged inter-fragment distance $b$ and inter-fragment angle $\theta$ for PEG$_{20}$ sampled from WTM and PICDiff. The error bars indicate the fluctuations of $b$ and $\theta$}
    \label{fig:peg-20}
\end{figure*}

We compared the structural ensembles generated by PICDiff with those from WTM simulations. First, $R_g$ distributions were compared (see Fig.~\ref{fig:peg-20}(b)), since $R_g$ is a commonly used feature for characterizing polymers. 
The $R_g$ distributions of both PEG$_{10}$ and PEG$_{20}$ generated by PICDiff agree well with those from WTM. Specifically, the peak locations of the $R_g$ distributions from PICDiff are slightly smaller than those obtained from physics-based simulations. The $R_g$ distribution of PEG$_{10}$ from PICDiff is slightly broader than that from WTM. Interestingly, the broadening of the $R_g$ distribution of PEG$_{20}$ from PICDiff agrees well with WTM. In addition to $R_g$, we also evaluated other statistics. We divided PEG$_{20}$ into 4 PEG$_{5}$ fragments. We investigated the interfragment distance and angle (Fig.~\ref{fig:peg-20}(a)). The resulting \(b\) and \(\theta\) distributions are compared with WTM simulations in Fig.~\ref{fig:peg-20}(c). Both distributions show good agreement with the WTM references. All these benchmarks suggest that PICDiff can accurately couple fragments to sample polymer conformations. 

\section{DISCUSSION}
An advantage of PICDiff is that it decomposes a system with a complex conformational space into subsystems with simpler conformational spaces. In the first example, the peptide–polymer complex is decomposed into a peptide and a polymer, and two SBDMs are used to represent them separately. In the second example, a long polymer chain is decomposed into fragments, with each SBDM modeling a single fragment. In general, the number of conformational states increases significantly with system size. Therefore, decomposing a large system into smaller systems can significantly reduce both the cost of sampling data required to train SBDMs and the cost of training SBDMs themselves. Moreover, the trained SBDMs can be reused in other applications. For example, the SBDM for $A\beta_{\mathrm{16-22}}$ can be used in PICDiff to model peptide binding to other polymers. 

We noticed that training SBDMs for PICDiff requires sampling diverse conformations, because $\uint$ may stabilize conformations that are high in free energy for isolated units. Therefore, a training dataset that covers a broad conformational space is needed. Our study demonstrates that different simulation trajectories can be combined to construct such a dataset, provided that the generated samples are properly reweighted. Interestingly, our previous studies suggest that transferable SBDMs may be able to sample protein conformations through guided inference, even when these conformations were not directly captured by the original SBDM \cite{wang2025extrapolating, liu2024exendiff}. Thus, PICDiff may be able to directly use existing foundation SBDMs or fine-tuned variants for the target system.

The modeling of $\uint$ is also crucial for PICDiff. The polymer-sampling example shows that errors in $\uint$ can bias the sampled ensemble distribution, and this bias becomes more significant when smaller units are used, because the inter-unit interface then accounts for a larger fraction of the system.
In principle, modeling $\uint$ should also account for changes in intra-unit energies caused by the surrounding environment. This issue is not severe in the two examples considered here. In the peptide–polymer example, polymer binding can alter the solvation environment and thereby change the inter-residue interactions within the peptide. Even without explicitly accounting for this effect, $\uint$ still leads to a reasonable ensemble distribution. In the polymer-sampling example, the averaged $b$ and $\theta$ values obtained from PICDiff agree with the WTM results, suggesting that the PEG$_5$ unit does not significantly contract or expand. This may be because PEG adopts relatively extended conformations, and neighboring fragments do not strongly perturb its local solvation structure.
Nevertheless, care is needed when constructing $\uint$, because environmental changes can, in certain cases, significantly affect the intra-unit potentials $U_1$ and $U_2$.

\section{CONCLUSIONS}
In this work, we introduced PICDiff, a general framework for coupling independently trained diffusion models through physics-based interactions to model complex molecular systems. This physics-guided coupling strategy extends the applicability of generative models by enabling the generation of configurations for multi-component systems and macromolecules, while reducing the need to collect structural data for these complex systems and train generative models directly on them.

We demonstrated PICDiff in two representative applications. For the A$\beta_{16-22}$-PEG$_{28}$ interface, PICDiff coupled the peptide and polymer diffusion models through nonbonded interactions and reproduced adsorption thermodynamics consistent with metadynamics-based reference calculations. For PEG chain construction, PICDiff assembled longer polymer chains from short fragment models by combining local bonded constraints with simplified nonbonded interactions. The resulting ensembles recovered key structural and thermodynamic properties, demonstrating that physically meaningful polymer statistics can be preserved through hierarchical fragment assembly.

At the same time, these results highlight several directions for future improvement. First, the accuracy of the coupling potential \(U_{\mathrm{int}}\) can be improved. 
Second, PICDiff can be integrated with more transferable molecular foundation models. 
More broadly, PICDiff provides a modular route for generative modeling of complex soft materials, biomolecular interfaces, and multicomponent assemblies. 

\begin{acknowledgments}
Y.W., Y.C., and M.C. gratefully acknowledge support from the National Science Foundation under EAR-2246687. 
\end{acknowledgments}

\section*{Data and Sofeware Availability}
The data that support the findings of this article will be openly available upon publication.
The software used for generating the data of this article is openly available\cite{picdiff_software}.


%

\bibliography{sample}

\end{document}